\documentstyle[epsf
              ,twoside
              ,csit
              ,times
              ]{article}

\newtheorem{df}{\sf Definition}[section]

\newtheorem{mth}{\sf Metatheorem}[section]

\title{The Set of Equations to Evaluate Objects}

\author{\large Larissa Ismailova \\[1.52mm]
\normalsize Vorotnikovsky per, 7, bld. 4 \\
\normalsize Institute for Contemporary Education ``JurInfoR-MSU'' \\
\normalsize Moscow, 103006, Russia \\
\normalsize {\tt lyui@jmsuice.msk.ru}
}

\institution{}

\begin{document}
\setcounter{page}{1}

\maketitle

\bibliographystyle{alpha}


\begin{abstract}
\noindent The notion of an {\em equational shell} is studied to
involve the objects and their environment. Appropriate {\em
methods} are studied as valid {\em embeddings} of refined objects.
The refinement process determines the linkages between the variety
of possible representations giving rise to variants of
computations. The case study is equipped with the adjusted
equational systems that validate the initial applicative
framework.
\end{abstract}



\section{Introduction}

One of the aims of this extended abstract is to
revisit the known ideas how to evalute the expression
using the applicative computational system.
Computation is sure to become one of the dominant
trend in computer research especially to carry out
object derivation processes.

{\em Objects}.
The remarks here can be taken as a suggestion to group
numerous aspects of `object' to result in a general computational
framework that gives a suitable scheme. This scheme can be useful
as a {\em primitive frame} to put important ideas of
evaluation
in a certain order.

{\em Representation}. Applications involve the excessively
complicated {\em representations} of objects that are equipped
with the {\em methods}.
An idea of object
generalizes the experimental or theoretical observations
concerning the behaviour of the selfcontained couples of data.
Among other representations the `data' is less of all
understood. Attempts to capture the most important features
of data lead to various mathematical ideas that are distant from
the selected model of computation
and result in a spectrum of far distant models.
The proposals here are to fill the gaps between them.

{\em Combinatory logic}. Combinators represent the basic properties
of functions, and combinatory logic represents a theory of functions
over {\em arbitrary} objects. The last notion
is significant to generate flexible data models.
Even more: combinatory logic is known as a sound amount of theoretical
and applied knowledge related to the ground nature of objects. It
supports: (1)~basic representation of arbitrary objects,
(2)~computational ideas with the objects,
(3)~integrity of both syntax and semantic.

Thus combinatory logic involves those entities
that combine both syntax
and semantical properties at the same time. This {\em was} known to
computer science theoreticians, but {\em was not} widely used in
applications at almost any scale.

Valid mathematical objects could be {\em embedded} into combinatory logic.
As embedding of an object is the main verification procedure within
combinatory logic then it is ready made for compiling an arbitrary
object into `computational instructions'
that in turn are combinators.

Those observations enforce the researcher to establish the regular
scheme to reconstruct all the vital
entities by the objects with useful mathematical properties.
In applications this is known as determining the {\em method}.

A brief outline of the refinement is observed as follows.
Imagine the fixed class of primary and derived objects
that is prescribed by the set of equations. The proposal is
to define the properties of the objects by adjusting the
initial set. The effect observed tends to capture more
meaning by the refined objects, and the process of refinement
evolves along distinct computational {\em methods}.

The outline of this extended abstract is as follows. The first and
second Sections contain a suitable formulation of a (higher order)
theory of functions. It is based on combinatory logic and the
relative computation theories referred as {\em shell}, or {\em
conceptual theory}.
The process of refinement is covered mainly in
the third Section. The connections of applicative and imperative
modes of computation are explicated. The refined objects are
embedded into the shell.
\section{Postulates}
To carry on with computation shell the minimal set
of equations has to be postulated. An easy start gives
the triple of primary objects ${\sf I,\ K,\ S}$
and the metaoperator of {\em application}. This
triple is power to maintain an {\em applicative
computation system} with the higher order functions.
\subsection{Applicative system}
 Let ${\sf I,\ K,\ S}$ to be the (mathematical) objects. Also an
infinite set of the indeterminants is added to support the supply
of {\em variables}. All the variables are included into the class
of objects. The objects determine the set of generic objects,
or atoms. The definition of derived objects is as follows
by induction on the complexity.
\begin{df}[Objects] (i) ${\sf I,\ K,\ S}$ and the variables are
the objects. (ii) If $a,\ b$ are the objects so is $(a\ b)$.
\end{df}
The step of induction needs the intuitive understanding. Thus,
the binary {\em application} operator $(\cdot\ \cdot)$ is taken
into game:
\begin{center}
$(\cdot\ \cdot): {\rm object}\ \times {\rm object}\ \rightarrow {\rm object}$.
\end{center}
It is the object generating operator that ranges over the objects.
A first object is viewed as the `function' while the second
is the `argument'. Hence the application operator enables function
to be applied to an argument that results in a generating of some
new object, or result of applying function to its argument,
and without application there is no chance to take a resulting value. \\
\indent To compare objects with other objects some (binary)
relation is to be defined. Usually this relation is referred as
the {\em conversion} and is determined by the {\em postulates} $(CL)$:
\[ ({\sf I})\ \ \ {\sf I}a=a,\ \ \ ({\sf K})\ \ \ {\sf K}ab=a,\ \ \
         ({\sf S})\ \ \ {\sf S}abc=ac(bc),
\]
\[
(\sigma)\ \ \  a=a,\ \ \ (\rho)\ \ \   \frac{a=b}{b=a},\ \ \
 (\tau)\ \ \ \frac{a=b,\ b=c}{a=c},
 \]
\[  (\mu)\ \ \ \frac{a=b}{ca=cb}, \qquad
(\nu)\ \ \   \frac{a=b}{ac=bc},
\]
where $a,\ b,\ c$ indicate the arbitrary objects and `=' is the
conversion relation.
\subsection{Alternative formulation}
Note that the class of objects above has the unique metaoperator,
namely application. For convenience the second metaoperator of
{\em abstraction} would be added:
\begin{center}
\begin{math}
(\lambda\cdot.\cdot):\ \ \ {\rm variable}\ \times {\rm object}\
 \rightarrow {\rm object}.
\end{math}
\end{center}
It is also object generating operator but it ranges over variables
and objects. After that the previous definition of an object may be
augmented by the additional step: {\it $(iii)$ If $x$ is a variable, $a$ is
an object then $\lambda x.a$, or $(\lambda x.a)$ is the object.} \\
\indent For convenience the agreement is added -- the left
associated parentheses may be demote (or remote) if needed.
The recent abstraction operator would be avoided.
\subsection{Basis}
To avoid the excessive objects the basis of disassembling
is needed. The following metatheorem validates the triple
${\sf I,\ K,\ S}$ to be the basis.
\begin{mth}[Disassembling] Any object $\lambda x.M$ may be
disassembled by case studying (according to induction on
complexity):
\[
\begin{array}{rl}
\textrm{(i)} & \lambda x.x={\sf I};\\
\textrm{(ii)}  & \lambda x.y = {\sf K}y, \ \ y \neq x;\\
\textrm{(iii)} & \lambda x.M'M''={\sf S}(\lambda x.M')(\lambda x.M'').
\end{array}
\]
\end{mth}
In fact, this metatheorem determines the primary basis.
\section{Creating a shell}
To verify the useful properties of basis ${\sf I,\ K,\ S}$ consider
an example of {\em embedding}. To be more rigorous add to postulates
$(CL)$ above the following schemes:
\[
\begin{array}{rl}
(\alpha) & \lambda x.a=\lambda y.[y/x]a,\ y \overline{\in} a
\ {\rm (congruency)};                 \\
(\beta)  & (\lambda x.a)b=[b/x]a\ {\rm (substitution)}; \\
(\xi)    & \displaystyle \frac{a=b}{\lambda x.a=\lambda x.b}; \\[0.5em]
(\eta)   & \lambda x.bx=b,\ x \overline{\in} b.
\end{array}
\]
(Note that $(\eta)$ determines $b$ as a concept.) The extended set
of postulates will be referred as $(CL\eta\xi)$.
\subsection{Restrictions}
The $(CL\eta\xi)$ formulation is given equationally, i.e. the
binary relation `$=$' of conversion can be specified as a kind
of equality. The additional equations seem to capture more
features of practically helpful objects. Up to the current stage
the consideration was purely syntactical. Now an attempt
to generate `the embedded applications' that essentially contain
semantics will be done.

For purely mathematical reasons the additional (and not
generic) combinators would simplify the notations.
Here some combinators are axiomatized by the following equations:
\[
\begin{array}{c}
{\sf I}x=x,\ {\sf C}xyz=xzy,\ {\sf B}xyz=x(yz), \\
{\sf K}xy=x,\ {\sf S}xyz=xz(yz),  \\
{\sf D}xy\equiv\lbrack x,y\rbrack\equiv\lambda r.rxy,
\ <f,g>\equiv \lambda t.[ft,gt], \\
{\sf \Phi}xyzw=x(yw)(zw),\ {\sf \Psi}xyzw=x(yz)(yw),  \\
{\sf B^2}\equiv{\sf BBB},\ {\sf C^2}xyzw=xwyz, \\
Curry \equiv \lambda h.\lambda xy.h[x,y],\ p[x,y]=x,\ q[x,y]=y.
\end{array}
\]
They will be used below to refine the properties of the initial
shell.
\subsection{Application}
Consider the set $(CL\eta\xi)$ of postulates with the additional
equation:
\[
\begin{array}{lcl}
{\sf B=\Psi(\Phi\ I)} &&\hspace{35mm} (\cdot(\cdot)=)
\end{array}
\]
The resulting set will be referred as $(CL\eta\xi)\ +\ (\cdot(\cdot)=)$.
To study the expressive power of this {\em conceptual equation} take
the indeterminants ${\sf V},\ M,\ N,\ \rho$  (possibly, variables,
or, at least, objects).

The left part application immediately gives:
\[ {\sf B V}MN\rho \stackrel{({\sf B})}{=}
             {\sf V}(MN)\rho \equiv \parallel MN \parallel\rho   \]
with the agreement ${\sf V}(\cdot )\equiv \|\cdot\|$, that
enables ${\sf V}$ as {\em evaluation} mapping.
The right part derivation results in
\[
\begin{array}{lcl}
{\sf \Psi(\Phi\ I)V}MN\rho &\stackrel{({\sf \Psi})}{=}&
              {\sf \Phi\ I(V}M)({\sf V}N)\rho \\
&\stackrel{({\sf \Phi})}{=}&{\sf I(V}M)\rho({\sf V}N\rho) \\
&\stackrel{({\sf I})}{=}&\|M\|\rho(\|N\|\rho)
\end{array}
\]
\noindent The direct observation gives the equation
\begin{center}
$\|MN\|\rho=\|M\|\rho(\|N\|\rho)$
\end{center}
that is implied by $(\cdot(\cdot)=)$.
\subsection{Ordered pair}
Consider the equation
\[
\begin{array}{lcl}
{\sf CB^2D=\Psi(\Phi\ D)} &&\hspace{35mm} ([\cdot,\cdot]=)
\end{array}
\]
\noindent in a context of $(CL\eta\xi)$, i.e. use the augmented shell
$(CL\eta\xi)\ +\ ([\cdot,\cdot]=)$. The left part concept
for ${\sf V},\ M,\ N,\ \rho$ generates the conversions as
follows:
\[
\begin{array}{lcl}
{\sf CB^2DV}MN\rho &\stackrel{({\sf C})}{=}&{\sf B^2VD}MN\rho \\
   &\stackrel{({\sf B^2})}{=}&{\sf V(D}MN)\rho   \\
   &\stackrel{({\sf D})}{=}&\|[M,N]\|\rho,
\end{array}
\]
\noindent and the right part gives:
\[
\begin{array}{lcl}
{\sf \Psi(\Phi\ D)V}MN\rho &\stackrel{({\sf \Psi})}{=}&
{\sf \Phi D (V}M)({\sf V}N)\rho \\
   &\stackrel{({\sf \Phi})}{=}&{\sf D}({\sf V}M\rho)({\sf V}N\rho)   \\
   &\stackrel{({\sf D})}{=}&[\|M\|\rho,\|N\|\rho].
\end{array}
\]
\noindent Thus the equation
\begin{center}
$\|[M,N]\|\rho=[\|M\|\rho,\|N\|\rho]$
\end{center}
is derived.
Discovering the conceptual equations $(\cdot(\cdot)=)$
and $([\cdot,\cdot]=)$, as may be shown below, refines the
properties of the initial shell $(CL)$ up to
{\em computational model} of general purpose.

For explicit
studying of $(CL\eta\xi)$ and
$(CL\eta\xi)~+~(\cdot(\cdot)=)~+~([\cdot,\cdot]=)$
computational properties the refined (and partially conversed)
consideration would be helpful. The concepts of main interest
are {\em constants} that gives rise to the {\em object constructor}.
\subsection{Constant object}

Often the formal systems involve the {\em constants}.
The notion or idea of a constant is assumed to be intuitively clear.
When the constants are viewed as the {\em relative entities} with respect to
some presupposed objects this idea is not so self-evident.
Let the {\em valuation} ${\sf V}$ and the {\em environment}
$\rho$ are selected
to be the point of relativization.
\begin{df}[constant object]
${\cal K}$ is defined to be the constant object relative to
the valuation ${\sf V}$ and the environment $\rho$
if and only if it is not
dependent on the valuation ${\sf V}$ and the environment $\rho$: \\
\[
\begin{array}{lr}
\parallel {\cal K}\parallel \rho \equiv {\sf V} {\cal K} \rho = {\cal K}& 
           \hspace{30mm} ({\cal K})
\end{array}
\]
\end{df}
Thus the equation 
(${\cal K}$) captures some important aspects and
{\em does} enrich our intuitive idea of a constant.
Moreover, provided ${\sf V}$ and $\rho$ are as above and ${\cal K}$ is
a constant object we have to assume for arbitrary object $x$:
\[
\begin{array}{lcl}
{\cal K}({\sf V}x\rho) &=& ({\sf V}{\cal K}\rho)({\sf V}x\rho) \hspace{20mm}
                                       {\rm\   by\ 
({\cal K})} \\
       &=& {\sf V}({\cal K}x)\rho.
\end{array}
\]
The last equation reflects a very natural principle that
`the valuation of application is the application of valuations'.
Similarly, one concludes:
\[
\begin{array}{lcl}
({\sf V}x\rho){\cal K} &=&({\sf V}x\rho)({\sf V}{\cal K}\rho) \hspace{20mm} {\rm\ by\ 
({\cal K}))}\\
       &=& {\sf V}(x{\cal K})\rho.
\end{array}
\]
The observations being accumulated result is the following
working rule: {\em the constant is extracted through the valuation
within some environment}. The importance of the equation 
(${\cal K}$)
erases a special {\em equational} principle of constant (${\cal K}$).
Actually, it would be better to construe the equation
${\sf V}{\cal K}\rho = {\sf C}{\sf V}\rho {\cal K}$ for
the combinator ${\sf C}$,
thus the principle  (${\cal K}$) would be reformulated as
${\sf C}{\sf V}\rho {\cal K} = {\cal K} = I{\cal K}$. \\
Let ${\cal K}$ be constructed as a variable by means of
$(CL\eta\xi)$. Then the equation (${\cal K}=$) is
derivable:
\[
\begin{array}{lr}
{\sf C}{\sf V}\rho = {\sf I} & \hspace{40mm} ({\cal K}=)
\end{array}
\]
This equation is intended in the desirable property of being a constant.
On the other hand using the equations
\[ {\sf V}{\cal K}\rho = {\cal K} = {\sf K}{\cal K}\rho , \]
and solving the equation (${\cal K}=$) for the evaluation ${\sf V}$
one obtains ${\sf V} = {\sf K}$.
The immediate consequence of this equation gives
\[ \parallel {\cal K}\parallel \rho = '{\cal K}\rho, \]
and hence $\parallel {\cal K}\parallel = '{\cal K}$ for $'={\sf K}$.
The symbol `\ $'$\ ' is the {\em quotation function} that
is analogous to the function {\em quote} in $LISP$.
For this solution of the equation 
(${\cal K}$) the following
conclusion is valid: \\
{\em the evaluation ${\sf K}$ gives the `constant' computational system,
i.e. evaluation views all the objects as ordinary constants.}
\subsection{Object constructor}
\subsubsection{Valuation of application}

Let $x,\ y$ be the objects evaluated as follows:\\
\[
\begin{array}{ll}
\parallel xy\parallel \rho
= \parallel (p\lbrack x,y\rbrack )
              (q\lbrack x,y\rbrack )\parallel \rho & \\
= \parallel {\sf S}pq\lbrack  x,y\rbrack \parallel\rho &    ({\sf S})\\
= {\sf S}pq(\parallel\lbrack x,y\rbrack\parallel\rho)  &        
({\cal K})\\
= {\sf S}pq\lbrack\parallel x\parallel\rho,
        \parallel y\parallel\rho \rbrack
                       &   (\lbrack\cdot ,\cdot\rbrack) \\
= (p\lbrack\parallel x\parallel \rho,
          \parallel y\parallel\rho \rbrack )
          (q\lbrack\parallel x\parallel \rho,\parallel y\parallel\rho
          \rbrack )  &         ({\sf S}) \\
= (\parallel x\parallel \rho)(\parallel y\parallel\rho) & (p,q)
\end{array}
\]
Here: ${\sf S}$ is a combinator, $p$ and $q$ are the first and second
projections respectively. The principles 
(${\cal K}$) and
the `valuation of pair' are used in this derivation.
Therefore the principle `valuation of application' is
derivable from the principles 
(${\cal K}$) and `valuation of pair'.
\subsubsection{Valuation of pair}
Let to analyze separately the derivation of principle
the `valuation of pair'.
The steps are analogous to those from the above:
\[
\begin{array}{lcll}
\parallel\lbrack x,y\rbrack\parallel\rho &\equiv&\parallel
{\sf D}xy\parallel\rho &\\
&=&\parallel {\sf D}x\parallel\rho(\parallel y\parallel\rho)& (\cdot (\cdot)) \\
&=& {\sf D}(\parallel x\parallel\rho)(\parallel y\parallel\rho) &  ({\cal K}) \\
&=& \lbrack\parallel x\parallel\rho,\parallel y\parallel\rho\rbrack
           & ({\sf D})
\end{array}
\]
Here: ${\sf D}$ is a pairing combinator. The principle `valuation of pair'
is derived from the`valuation of application' and 
(${\cal K}$).
Hence the principle `valuation of pair' is derivable from
the principles 
(${\cal K}$) and `valuation of application'.

\subsubsection{Redundancy of computational principles}

As was observed above the principles 
(${\cal K}$), $\cdot(\cdot)$ and
$\lbrack\cdot,\cdot\rbrack$ are mutually dependent.
Thus some redundant entities would be eliminated. The possible
postulates are the principles as follows:\\
\indent (1) $\parallel {\cal K} \parallel\rho = {\cal K}$; \\
\indent (2) either `valuation of application' or `valuation of pair'.

\section{Equational notation}

Now let apply the computational principles to the combinators.
Suppose ${\sf V}, M, N, \rho$ are the variables.
\[
\begin{array}{lcl}
\parallel MN\parallel\rho &=& (\parallel M\parallel\rho)
                              (\parallel N\parallel\rho) \\
      &=& {\sf V}(MN)\rho = ({\sf V}M\rho)({\sf V}N\rho)  \\
      &=& {\sf BV}MN\rho = {\sf \Phi I}({\sf V}M)({\sf V}N)\rho  \\
      &=& {\sf\Psi} ({\sf \Phi I}){\sf V}MN\rho.
\end{array}
\]
From the equation ${\sf BV}MN\rho = {\sf\Psi}({\sf\Phi I})
{\sf V}MN\rho$ given above
the characteristic equation $(\cdot(\cdot)=)$ is derivable:
\[ {\sf B} = {\sf\Psi}({\sf\Phi I}) \hspace{30mm} (\cdot(\cdot)=) \]
This equation is understood as the {\em equational} notation
for the principle `evaluation of application' whereas
${\sf V}$ is the valuation, $M,N$ are the objects, and $\rho$ is
the environment or assignment. \\
The same reasons are applied to the equational notation of the
`evaluation of pair': \\
\[
\begin{array}{lcl}
\parallel \lbrack M,N\rbrack\parallel\rho
   &=& \lbrack\parallel M\parallel\rho,\parallel N\parallel\rho\rbrack \\
   &=& {\sf V}\lbrack M,N\rbrack\rho = \lbrack {\sf V}M\rho,
{\sf V}N\rho\rbrack  \\
   &=& Curry\ {\sf V}MN\rho = {\sf D}({\sf V}M\rho)({\sf V}N\rho)  \\
   &=& {\sf\Phi D}({\sf V}M)({\sf V}N)\rho   \\
   &=& {\sf\Psi}({\sf\Phi D}){\sf V}MN\rho
\end{array}
\]
The immediate consequence is the equation $(\lbrack\cdot ,\cdot\rbrack =)$: \\
\[ Curry = {\sf \Psi(\Phi D)} \hspace{30mm} (\lbrack\cdot ,\cdot\rbrack =)\]
The modified equation takes into account $Curry = {\sf CB^2D}$.
Thus
\[ {\sf CB^2D = \Psi(\Phi D)} \hspace{30mm} (\lbrack\cdot ,\cdot\rbrack =)\]

\subsection{Modified equation to evaluate the application}

The following observation would be fruitful for further derivations.
The evaluation of $\parallel x y\parallel\rho$ is likely to involve
the definition of $\varepsilon$.
From $xy = \varepsilon \lbrack x,y\rbrack$ the following equations
are valid:
\[
\begin{array}{lcll}
\parallel xy\parallel\rho
 &=& \parallel\varepsilon\lbrack x,y\rbrack\parallel\rho &  \\
 &=&\varepsilon(\parallel\lbrack x,y\rbrack\parallel\rho) & \\
 &=&\varepsilon\lbrack\parallel x\parallel\rho,
     \parallel y\parallel\rho\rbrack  & (by\ \lbrack\cdot,\cdot\rbrack ) \\
 &=&(\parallel x\parallel\rho)(\parallel y\parallel\rho)
                 & (by\ \varepsilon)
\end{array}
\]

\subsection{Currying, application and product}

Let $z$ be equal to the ordered pair i.e. $z=\lbrack u,v\rbrack$. Of course,
from the equations $u=pz$ and $v=qz$ we derive
$z=\lbrack pz,qz\rbrack =<p,q>z$. Having in mind the equation $z={\sf I}z$
and ignoring the {\em type} considerations it is easy to show:
\[ <p,q> = {\sf I} \hspace{30mm} (\times=) \]
Suppose $h=\varepsilon$ in the definition
$h\lbrack x,y\rbrack = Curry\ h\ xy$. The immediate consequence is the
following:
\[ \varepsilon\lbrack x,y\rbrack = xy = Curry\ \varepsilon\ xy \]
Adding the equation $xy={\sf I}\ xy$ it is easy to show:
\[ Curry\ \varepsilon = {\sf I} \]
The equation above interconnects the currying $Curry$ and the
explicit application $\varepsilon$.
The following is derivable from the equation $(\times =)$:
\[
\begin{array}{lcll}
hz &=& h\lbrack pz,qz\rbrack & \\
   &=& Curry\ h(pz)(qz)       & (by\ (\times =) ) \\
   &=& (Curry\ h\circ p)z(qz) & (by\ \circ)  \\
   &=& \varepsilon \lbrack(Curry\ h\circ p)z,qz\rbrack & (by\ \varepsilon)\\
   &=& (\varepsilon\circ<Curry\ h\circ p,q>)z,  &  (by\ <\cdot,\cdot> )
\end{array}
\]
For arbitrary variable $z$ in the equations above one concludes:
\[ h = \varepsilon\circ <Curry\ h\circ p,q>
           \hspace{30mm} (\lbrack\cdot,\cdot\rbrack) \]
The last equation gives characteristics of the computations
with the {\em ordered pairs}. \\
The modified derivation gives the following:
\[
\begin{array}{lcl}
kxy &=&\varepsilon\lbrack kx,y\rbrack \\
    &=&(\varepsilon\circ<k\circ p,q>)\lbrack x,y\rbrack \\
    &=& Curry(\varepsilon\circ<k\circ p,q>)xy
\end{array}
\]
The derivation above generates the equation $(\cdot(\cdot))$:
\[ k = Curry(\varepsilon\circ<k\circ p,q>) \hspace{30mm}  (\cdot(\cdot)) \]
\noindent that characterizes the computations with the {\em applications}.
Combinators and combinatory logics produce some additional entities
e.g. {\em product} and {\em coproduct}.

\section*{Conclusions}
\addcontentsline{toc}{section}{Conclusions}
Main results are briefly summarized as follows.

1. Varying with different researches the nature of `object' from
a computational point of view would be captured,
represented and embedded into a kind of
primitive frame. This scheme operates within a theory of functions
concerning {\em combinatory logic} and generates
a primary conceptual shell.

2. Combinators give a sound {\em substrate} to produce a data object
model. The objects in use inherit both syntax and semantics of the
initial idea of object. This leads to and object-as-functor
computations and generates a refinement process to capture
the {\em methods} for individual objects.

3. It could be shown that the concepts are embedded into the shell
and inherit the logical properties of the objects. The higher order
theory (with {em the descriptions}) is in use.

4. The refinement process suits the equational conditions.
The distinct methods are to be studied within an equational
framework.



\end{document}